%% file: manuscript.tex
\documentclass[a4paper,fleqn,usenatbib]{mnras}

\usepackage{mathptmx}

\usepackage[T1]{fontenc}
\usepackage{ae,aecompl}

\usepackage{graphicx}	
\usepackage{amsmath}	
\usepackage{amssymb}	
\usepackage{comment}
\usepackage[%
        ]{hyperref}

\input{shortcuts.tex}

\title[
Efficient, uninformative sampling of LDCs
]{
Efficient, uninformative sampling of limb-darkening coefficients for a 
three-parameter law
}

\author[D. M. Kipping]{
David M. Kipping$^{1}$\thanks{E-mail:\href{mailto:\myemail}{\myemail}}
\\
$^{1}$Department of Astronomy, Columbia University, 550 W 120th St, New York, 
NY 10027, USA
}

\date{Accepted 2015 October 12. Received 2015 October 8; in original form 2015 September 8}

\pubyear{2015}

\begin{document}
\label{firstpage}
\pagerange{\pageref{firstpage}--\pageref{lastpage}}
\maketitle

\begin{abstract}
Stellar limb-darkening impacts a wide range of astronomical measurements. The
accuracy to which it is modelled limits the accuracy in any covariant parameters 
of interest, such as the radius of a transiting planet. With the ever growing 
availability of precise observations and the importance of robust estimates of 
astrophysical parameters, an emerging trend has been to freely fit the
limb-darkening coefficients (LDCs) describing a limb-darkening law of choice, in
order to propagate our ignorance of the true intensity profile. In practice,
this approach has been limited to two-parameter limb-darkening laws, such as the
quadratic law, due to the relative ease of sampling the physically allowed
range of LDCs. Here, we provide a highly efficient method for sampling LDCs
describing a more accurate three-parameter non-linear law. We first derive 
analytic criteria which can quickly test if a set of LDCs are physical, although 
naive sampling with these criteria leads to an acceptance rate less than 1\%. We 
then show that the loci of allowed LDCs can be transformed into a cone-like 
volume, from which we are able to draw uniform samples. We show that samples 
drawn uniformly from the conal region are physically valid in 97.3\% of 
realizations and encompass 94.4\% of the volume of allowed parameter space. We 
provide \python\ and \fortran\ code (\LDC) to sample from this region (and 
perform the reverse calculation) at \link, which also includes a subroutine to 
efficiently test whether a sample is physically valid or not.
\end{abstract}




\section{INTRODUCTION}
\label{sec:intro}

\input{introduction.tex}

\section{PHYSICAL CRITERIA}
\label{sec:criteria}

\input{criteria.tex}

\section{BOUNDS ON THE LDC\tiny{s}}
\label{sec:bounds}

\input{bounds.tex}

\section{NUMERICAL TESTING OF THE CRITERIA}
\label{sec:criteria_tests}

\input{criteria_tests.tex}

\section{TRANSFORMATIVE GEOMETRY}
\label{sec:transformations}

\input{transformations.tex}

\section{DISCUSSION}
\label{sec:discussion}

\input{discussion.tex}

\section*{Acknowledgements}

DMK acknowledges support from the CfA Menzel fellowship programme and thanks
the anonymous reviewer for their helpful comments.

\bsp	
\label{lastpage}
\end{document}

%% file: shortcuts.tex
\newcommand{\www}{https://github.com/davidkipping/LDC3}
\newcommand{\link}{\href{\www}{this http}}

\newcommand{\myemail}{d.kipping@columbia.edu}
\newcommand{\LDC}{\textsc{ldc3}}
\newcommand{\python}{\textsc{python}} 
\newcommand{\fortran}{\textsc{fortran}} 
\newcommand{\I}{\textbf{I}}
\newcommand{\II}{\textbf{II}}
\newcommand{\III}{\textbf{III}}
\newcommand{\boldalpha}{\boldsymbol{\alpha}}

%% file: introduction.tex

Stellar limb-darkening affects a wide range of astronomical observations, such
as exoplanetary transits \citep{mandel:2002}, microlensed light curves 
(e.g. \citealt{witt:1995}; \citealt{zub:2011}), rotational modulations 
\citep{macula:2012}, ellipsoidal variations \citep{morris:1993}, interferometric 
images of stars (e.g. \citealt{aufdenberg:1995}) and eclipsing binaries 
\citep{kopal:1950}. When interpreting such observations, the assumed shape of 
the limb-darkening intensity profile may significantly affect the inferred 
parameters of interest \citep{csizmadia:2013}. Consequently, an accurate 
treatment of limb-darkening is crucial, even when the effect itself is a 
``nuisance'' phenomenon.

In many practical cases, limb-darkening is treated by describing the stellar 
intensity profile with a closed-form analytic model. This model is usually 
called the ``limb-darkening law'', which is designed to provide an accurate 
analytic approximation the true profile. The major advantage of modelling the
intensity profile analytically is that the astronomical phenomena under
investigation may also be modelled analytically, offering computational
expedience and greater physical insight. As an example, in the field of 
exoplanet transits, the shape of a transit light curve can be described with 
a closed-form, analytic model under the assumption of a polynomial-based 
description of the stellar intensity profile \citep{mandel:2002,gimenez:2006}.

All of the commonly used limb-darkening laws may be described as a linear sum of
one or more simple functions, each of which is weighted by a so-called 
limb-darkening coefficient (LDC). For example, the popular quadratic 
limb-darkening law describes the intensity profile as a quadratic series 
expansion with respect to the angle between the line of sight and the emergent 
intensity ($\mu$) and has two LDCs \citep{kopal:1950}. These LDCs control the 
shape of the intensity profile, subject to the flexibility granted by the 
limb-darkening law's complexity. It is therefore necessary to make at least two 
major decisions with how to treat limb-darkening; what limb-darkening law should 
be used and what LDCs should be assigned to this law?

A typical approach for assigning LDCs is to adopt a set of so-called 
``theoretical'' LDCs. In this framework, one first simulates an intensity 
profile using a sophisticated stellar atmosphere model at a particular 
wavelength or over a chosen integrated bandpass. One then takes this 
simulation and regresses the limb-darkening law of choice to it by finding
the maximum likelihood set of LDCs (in some instances the LDCs are restricted
to ensure physically sound profiles). Since the simulated profile is sensitive 
to several parameters defining the stellar surface (e.g. effective temperature, 
metallicity, surface gravity), several groups have produced grid tabulations of 
LDCs for a wide range of plausible inputs and assumed limb-darkening law (e.g.
see \citealt{vanhamme:1993}, \citealt{diaz:1995}, \citealt{claret:2000} and
\citealt{sing:2010}).

In recent years, there has been a shift away from using theoretical LDCs in 
favour of freely fitting LDCs simultaneous to the exploration of the other 
parameters of interest (e.g. \citealt{knutson:2007}, \citealt{kipbak:2011a,
kipbak:2011b} and \citealt{kreidberg:2014}). This approach, enabled by advances 
in modern computers, allows one to propagate our ignorance regarding the shape 
of the intensity profile into the estimation of the parameters of interest, 
leading to more accurate estimates parameter uncertainties. Such a procedure
also allows one to decouple from any possible errors in the stellar atmosphere
models themselves, which have been shown to sometimes be inconsistent with
observed limb-darkening profiles \citep{howarth:2011,epinoza:2015}.

One major challenge when attempting to freely fit LDCs is that many combinations
of the LDCs are unphysical. For example, a specific choice of LDCs may result
in the intensity profile being occasionally negative. These regions of 
unphysical parameter space may coincide with likelihood minima, leading to
erroneous parameter posterior distributions. Softer, extended likelihood
minima (typical of low signal-to-noise data) may spill over into the unphysical
parameter space, causing the final parameter posteriors to be marginalized over 
large swaths of unphysical parameter space leading to unnecessarily swollen 
error bars.

There are two solutions to this problem. The first is to define a set of 
criteria which quickly allow us to identify forbidden combinations of the LDCs.
Armed with such criteria, one could then test each realization and accept or
reject the realization accordingly. This works fine unless the volume of
forbidden parameter space is large, in which case one will severely impede
a regression algorithm's efficiency since most trials are being rejected
as unphysical. A more elegant solution is to directly sample exclusively (or
at least efficiently) from the physically allowed volume of parameter space (and 
also in such a way that the LDCs can be uniformly, or nearly uniformly, 
sampled). This provides the benefits of a very high efficiency, physically sound 
priors and no need for testing criteria at each realization. However, this 
approach comes at the cost of the initial intellectual investment to actually 
solve how to perform efficient sampling in the first place.

Because direct sampling of physical LDCs is challenging, the most complex 
limb-darkening law for which this feat has yet been achieved is the simple 
quadratic law, where the intensity profile of the star is given by

\begin{align}
I(\mu)/I(1) &= 1 - u_1 (1-\mu) - u_2 (1-\mu)^2,
\label{eqn:Iquad}
\end{align}

where $I(1)$ is the specific intensity at the centre of the disc, $u_1$ and
$u_2$ are the quadratic LDCs and $\mu$ is the cosine of the angle between the 
line of sight and the emergent intensity. In \citet{LD:2013}, we showed that
physically sound $u_1$ and $u_2$ LDCs reside within a triangle on the
$u_1$-$u_2$ plane, from which uniform samples can be easily drawn by
re-parameterizing to $q_1 = (u_1+u_2)^2$ and $q_2 = 0.5u_1(u_1+u_2)^{-1}$.
Other simple two-parameter laws were considered as well in this work.

Despite these successes with two-parameter laws, these laws are fundamentally
limited in their ability to accurately describe realistic limb-darkening 
profiles. This translates to systematic uncertainty in any and all model 
parameters which are covariant with the limb-darkening properties. This may,
for example, limit our ability to accurately measure the radius of a transiting
exoplanet and thus make inferences of its composition. We are therefore motivated
to extend the \citet{LD:2013} analysis to a more sophisticated limb-darkening
law.

The most accurate closed-form limb-darkening law is the \citet{claret:2000} 
four-parameter non-linear law, described by

\begin{align}
I(\mu)/I(1) = 1 &- c_1 (1-\mu^{1/2}) - c_2 (1-\mu) \nonumber\\
\qquad&- c_3 (1-\mu^{3/2}) - c_4 (1-\mu^{2}).
\label{eqn:Iclaret}
\end{align}

In numerous independent studies, this law has been found to provide the most 
accurate description of simulated intensity profiles (e.g. 
\citealt{claret:2000}, \citealt{sing:2010} and \citealt{magic:2015}) versus 
competing models. This is, however, not surprising since this law also utilizes 
the greatest number of LDCs. Analytically identifying the unphysical 
combinations of these four LDCs remains an outstanding and formidable challenge. 
A more tractable problem that we consider in this work is the allowed volume of 
LDCs (and methods to directly sample from said volume) in the case of the 
\citet{sing:2009} three-parameter law:

\begin{align}
I(\mu)/I(1) &= 1 - c_2 (1-\mu) - c_3 (1-\mu^{3/2}) - c_4 (1-\mu^{2}).
\label{eqn:Ising}
\end{align}

\citet{sing:2010} argues that dropping the $c_1$ term is motivated by
solar data \citep{neckel:1994} and 3D stellar models \citep{bigot:2006}, which
show that $I(\mu)$ varies smoothly at small $\mu$, meaning that a $\mu^{1/2}$
term is superfluous. We therefore argue that the \citet{sing:2009} law
offers both a significant improvement in accuracy and yet is simple enough for
us to analytically constrain the allowed LDCs.

%% file: criteria.tex

\subsection{Physical Conditions}
\label{sub:physicalconditions}

We define the following \textit{physical conditions} with respect to a limb
darkened stellar intensity profile:

\begin{itemize}
\item[{\textbf{(I)}}] an everywhere-positive intensity profile,
\item[{\textbf{(II)}}] a monotonically decreasing intensity profile from the
centre of the star to the limb,
\item[{\textbf{(III)}}] the intensity profile has a negative curl at the limb.
\end{itemize}

Physical conditions \I\ and \II\ are the same two imposed in our previous paper,
\citet{LD:2013}. As noted in that work, limb brightening is possible for 
narrow-band observations (e.g. see \citealt{schlawin:2010}) and such behaviour
is not considered in this work either. Physical condition \III\ is motivated by 
the expectation that the intensity rapidly drops off towards the limb (and is 
discussed in more detail later in \S\ref{sub:criterionE}). Throughout this work,
we refer to a set of LDCs satisfying these three conditions as being physical, 
and LDCs otherwise are defined as unphysical. In what follows, we explore the 
consequences of these simple constraints.

\subsection{Criterion A}

Physical condition \I\ demands that $I(\mu)>0$ $\forall$ $0\leq\mu<1$. We begin 
by evaluating this condition at two extrema of $\mu\to1$ and $\mu\to0$, in a 
similar manner to the approach adopted in \citet{LD:2013}:

\begin{align}
\lim_{\mu\to1} I &= 1 > 0,\nonumber \\
\lim_{\mu\to0} I &= 1 - c_2 - c_3 -  c_4 > 0.
\end{align}

The upper line clearly has no constraining power, but the second line provides
our first criterion of

\begin{align}
c_2 + c_3 + c_4 < 1.
\label{eqn:criterionA}
\end{align}

\subsection{Criteria B and C}

Next, we consider physical condition \II, which demands that 
$\partial I/\partial \mu>0$ $\forall$ $0\leq\mu<1$:

\begin{align}
\frac{ \partial I(\mu) }{\partial \mu } &= c_2 + \frac{3}{2} c_3 \mu^{1/2} + 2 c_4 \mu.
\end{align}

As was done in the previous subsection, let us evaluate the above in the extreme 
cases of $\mu\to1$ and $\mu\to0$, yielding

\begin{align}
\lim_{\mu\to1} \frac{ \partial I(\mu) }{\partial \mu } &= c_2 + \frac{3}{2} c_3 + 2 c_4 > 0,\nonumber\\
\lim_{\mu\to0} \frac{ \partial I(\mu) }{\partial \mu } &= c_2 > 0.
\end{align}

These two expressions provide our criteria B and C, which are respectively
given by

\begin{align}
2 c_2 + 3 c_3 + 4 c_4 > 0,
\label{eqn:criterionB}
\end{align}

and

\begin{align}
c_2 > 0.
\label{eqn:criterionC}
\end{align}

\subsection{Criterion D}

Physical condition \II\ tells us that the intensity decreases from the centre of 
the star to the limb. This implies that the intensity everywhere (except at the
centre of the star) is less than that present at the centre of the star, or
mathematically that

\begin{align}
I(\mu) < \lim_{\mu\to1} I(\mu).
\end{align}

One simple closed-form result from this constraint occurs by comparing the
intensity at the limb to the centre via:

\begin{align}
\lim_{\mu\to0} I(\mu) < \lim_{\mu\to1} I(\mu).
\end{align}

This condition, derived by physical condition \II, provides our fourth
criterion,

\begin{align}
c_2 + c_3 + c_4 > 0.
\label{eqn:criterionD}
\end{align}

\subsection{Criterion E}
\label{sub:criterionE}

Consider the behaviour of the intensity profile at the limb. By virtue of
condition \II, the intensity profile must be decreasing as we approach the 
boundary. We therefore expect a negative gradient with respect to $r$, or 
equivalently a positive gradient with respect to $\mu$, since 
$\partial r/\partial \mu$ is always negative. We also expect that at the limb 
the gradient of the gradient (i.e. the curl) is negative. This is consistent 
with the asymptotic-like behaviour expected due to foreshortening near the limb, 
causing the gradient to become ever-more negative and defines physical condition 
\III. Note that a negative curl with respect to $r$ is equivalent to a negative 
curl with respect to $\mu$, since we now multiply by $(\partial r/\partial \mu)$ 
twice, leading to a double negative. The curl may be expressed as

\begin{align}
\frac{\partial^2 I(\mu)}{\partial \mu^2} = \frac{3}{4} c_3 + 2 c_4 \mu^{1/2}.
\end{align}

At the limb then ($\mu \to 0$), we expect that

\begin{align}
\lim_{\mu\to0} \Big(\frac{\partial^2 I(\mu)}{\partial \mu^2}\Big) < 0,
\end{align}

which defines criterion E,

\begin{align}
c_3 < 0.
\label{eqn:criterionE}
\end{align}

\subsection{Criterion F}
\label{sub:criterionF}

Physical condition \I\ requires that $I(\mu)$ is everywhere positive. Combining
\I\ with \II\ implies that $I(\mu)$ must be less than unity everywhere, which
one may consider to be physical condition \textbf{I'}. Writing this out along 
with \II, one may show that

\begin{align}
-c_2 \mu - c_3 \mu^{3/2} - c_4 \mu^2 > -c_2 - c_3 - c_4,\\
\frac{2}{3} c_2 \mu + c_3 \mu^{3/2} + \frac{4}{3} c_4 \mu^2 > 0.
\end{align}

Adding the two inequalities shown above cancels out the $c_3 \mu^{3/2}$ terms
and leaves us with a quadratic equation:

\begin{align}
c_4 \mu^2 - c_2 \mu > -3(c_2 + c_3 + c_4).
\end{align}

From Criterion A, we know that the sum of the coefficients must be less than 
unity, implying that in the limit of $\mu\to1$, we have,

\begin{align}
c_4 - c_2 > -3.
\label{eqn:c4c2diff}
\end{align}

Starting from criterion B and invoking criterion E, we can also show that:

\begin{align}
2c_2 + 3c_3 + 4c_4 > 0,\nonumber\\
3c_3 > -2c_2 - 4c_4,\nonumber\\
-2c_2 - 4c_4 < 3c_3 < 0,\nonumber\\
c_2 + 2c_4 > 0.
\label{eqn:c4c2sum}
\end{align}

Summing Equations~\ref{eqn:c4c2diff} \& \ref{eqn:c4c2sum} together yields

\begin{align}
c_4 > -1.
\end{align}

Through numerical experimentation, we find that applying the slightly more 
conservative bound of $c_4>0$ yields a more symmetric loci of allowed points
(as discussed later in \S\ref{sec:transformations}), from which it is easier to 
directly sample. We therefore modify criterion F to,

\begin{align}
c_4 > 0.
\label{eqn:criterionF}
\end{align}

\subsection{Criterion G}
\label{sub:criterionG}

For our final criterion, we begin by considering physical condition \II:

\begin{align}
\frac{ \partial I(\mu) }{\partial \mu } &> 0,\nonumber\\
2 c_4 \mu + \frac{3}{2} c_3 \mu^{1/2} + c_2 &> 0.
\end{align}

The gradient expressed must be everywhere-positive and so let us compute
the minimum gradient possible, which occurs when the curl equals zero, or
when

\begin{align}
\frac{\partial (2 c_4 \mu + \frac{3}{2} c_3 \mu^{1/2} + c_2) }{ \partial \mu } &= 0,\nonumber\\
\frac{3}{4} c_3 \mu^{-1/2} + 2 c_4 &= 0.
\end{align}

Therefore, the minimum gradient occurs when $\mu=\mu_{\mathrm{min}}$, where we 
define

\begin{align}
\mu_{\mathrm{min}}^{1/2} = -\frac{3c_3}{8c_4}.
\end{align}

In the case where criterion E and F are in effect, then $c_3$ is negative and 
$c_4$ is positive meaning that $\mu_{\mathrm{min}}$ is a real number.

The point $\mu_{\mathrm{min}}$ may or may not be within the range $0<\mu<1$. If 
indeed it is, then implicitly $-1 < 3 c_3/(8 c_4) < 0$ and we require that 
the gradient at this point is positive, giving

\begin{align}
\mathrm{if}\,\,\,-1 < \frac{3 c_3}{8 c_4} < 0\,\mathrm{\,\,\,then}\nonumber\\
c_2 > \frac{9 c_3^2}{32 c_4}.
\end{align}

As with criterion F, we find through numerical tests in 
\S\ref{sec:transformations} that the loci of points can be made symmetric if we 
impose criterion G under \textit{all} circumstances, not just when 
$-1 < 3 c_3/(8 c_4) < 0$. We therefore modify criterion G to

\begin{align}
c_2 > \frac{9 c_3^2}{32 c_4}.
\label{eqn:criterionG}
\end{align}

\subsection{Summary of Analytic Criteria}
\label{sub:summaryofcriteria}

To summarize, our seven analytic criteria on the three LDCs are

\begin{align}
c_2 + c_3 + c_4 &< 1,\,\,\,\mathbf{[A]}\nonumber\\
2 c_2 + 3 c_3 + 4 c_4 &> 0,\,\,\,\mathbf{[B]}\nonumber\\
c_2 &> 0, \,\,\,\mathbf{[C]}\nonumber\\
c_2 + c_3 + c_4 &> 0, \,\,\,\mathbf{[D]}\nonumber\\
c_3 &< 0, \,\,\,\mathbf{[E]}\nonumber\\
c_4 &> 0, \,\,\,\mathbf{[F]}\nonumber\\
c_2 &> \frac{9 c_3^2}{32 c_4}. \,\,\,\mathbf{[G]}
\label{eqn:ALLcriteria}
\end{align}

Using the three physical conditions only, we note that criterion F should
strictly be $c_4>-1$. We have modified this criterion to be slightly more 
conservative so that the loci of allowed LDCs can be transformed into a 
symmetric cone shape depicted later in \S\ref{sec:transformations}. Similarly, 
criterion G strictly only applies when $-1 < 3 c_3/(8 c_4) < 0$ if one uses 
the three physical conditions. We again modify the criterion such that it 
applies under all circumstances, in order to yield a more symmetric loci, as
shown later in \S\ref{sec:transformations}. We provide \python\ and \fortran\
code (\LDC) to test whether these criteria hold, for which the user can also use
the unmodified versions of the criteria if desired (available at \link).

In \S\ref{sec:criteria_tests}, we explore the consequences of these 
modifications and perform numerical tests demonstrating the effectiveness of the 
seven criteria. First though, we calculate the allowed maxima/minima on each
LDC using the seven criteria, as shown in the next section, \S\ref{sec:bounds}.

%% file: bounds.tex

In the cases of criteria C, E and F, we have derived a hard boundary on the 
isolated LDCs; for example, $c_2>0$ from criterion C. However, it is also useful 
to know about the other extrema; for example, what is the maximum bound on 
$c_2$? Working in the $\{c_2,c_3,c_4\}$ parameter space, which we refer 
to as the ``original'' reference frame, these bounds define the smallest cuboid 
within which the allowed loci reside.

\subsection{$c_3$-$c_2$ plane}
\label{sub:c3c2plane}

Starting from the seven criteria, how can we calculate the limits on each LDC?
We choose to proceed by means of numerically-guided analytic reasoning via
inspection of the projected 2D planes. 

\begin{figure}
\includegraphics[width=\columnwidth]{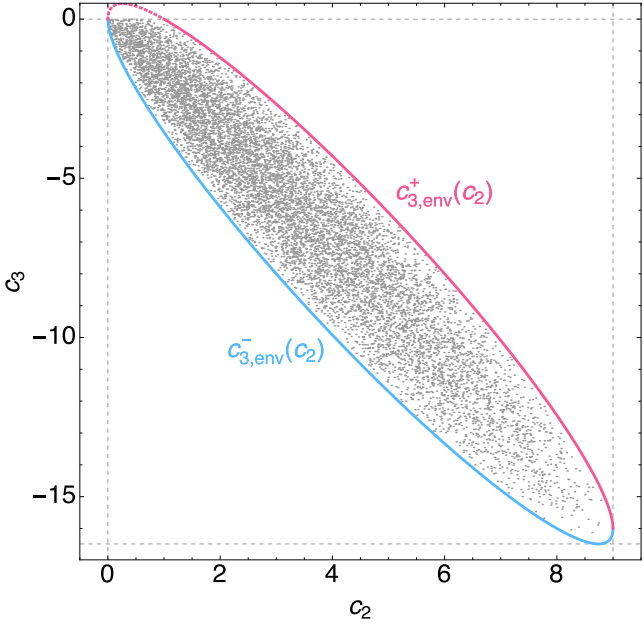}
\caption{
Loci of points satisfying criteria A-G, plotted in the $c_3$-$c_2$ plane.
The pink/blue lines are given by $c_{3,\mathrm{env}}^{\pm}$.
}
\label{fig:c3c2surface}
\end{figure}

We begin by generating a large number of random points drawn uniformly in 
$\{c_2,c_3,c_4\}$ and then reject any point which does not satisfy criteria A-G. 
The surviving points are then saved. We plot the loci of these points on the 
$c_3$-$c_2$ plane in Fig.~\ref{fig:c3c2surface}.

Inspection of Fig.~\ref{fig:c3c2surface} reveals that the loci of points are
nearly perfectly enveloped by the functions $c_3=(4/9) (-4c_2 \pm \sqrt{2 c_2} 
\sqrt{9-c_2})$, with the exception of a small region prohibited by criterion E
($c_3<0$). The envelope function may be derived starting from criteria A and G,
as follows:

\begin{align}
c_4 &< 1 - c_3 - c_2,\,\,\mathbf{[A]}\nonumber\\
c_4 &> \frac{9 c_3^2}{32 c_2},\,\,\mathbf{[G]}\nonumber\\
\implies \frac{9 c_3^2}{32 c_2} &< 1 - c_3 - c_2.
\end{align}

which may be re-arranged to

\begin{align}
9 &\Big( c_3 - \frac{4}{9} \big(-4c_2 - \sqrt{2 c_2} \sqrt{9-c_2}\big) \Big) \nonumber\\
\times &\Big( c_3 - \frac{4}{9} \big(-4c_2 + \sqrt{2 c_2} \sqrt{9-c_2}\big) \Big) < 0
\end{align}

For the above to hold, then $c_3$ must be larger than one of the radicals but
always less than the other, giving two possible sets of envelope functions.
We define the first one as:

\begin{align}
c_3 > c_{3,\mathrm{env}}^{-}(c_2) = \frac{4}{9} (-4c_2 - \sqrt{2 c_2} \sqrt{9-c_2}), \\
c_3 < c_{3,\mathrm{env}}^{+}(c_2) = \frac{4}{9} (-4c_2 + \sqrt{2 c_2} \sqrt{9-c_2}).
\end{align}

In Fig.~\ref{fig:c3c2surface}, the pink line depicts 
$c_{3,\mathrm{env}}^{+}(c_2)$, whilst the blue line depicts 
$c_{3,\mathrm{env}}^{-}(c_2)$. This demonstrates that indeed our first guess for
the form of the solution is correct. In principal though, we note that an 
alternative solution is $c_{3,\mathrm{env}}^{+}(c_2)<c_3<
c_{3,\mathrm{env}}^{-}(c_2)$.

From Fig.~\ref{fig:c3c2surface}, one can see that there is a minimum allowed
$c_3$ value and a maximum $c_2$. The maximum $c_2$ coincides where the two
envelope functions meet, meaning that $\sqrt{2 c_2} \sqrt{9-c_2}=0$, giving
$c_2=9$. The minimum $c_3$ value can be found by minimizing the lower envelope,
which occurs at $c_2=(3/2)(3+2\sqrt{2})$, corresponding to the minimum value of
$c_3=-8-6\sqrt{2}$, or $-16.4853...$. This therefore provides two of the missing
three LDC bounds we seek.

\subsection{$c_4$-$c_2$ plane}
\label{sub:c4c2plane}

We may repeat this exercise in the $c_4$-$c_2$ plane, and the resulting loci
are shown in Fig.~\ref{fig:c4c2surface}. As before, two lines appear to 
nearly perfectly envelope the loci of points, which can be derived starting from
criteria A and G:

\begin{figure}
\includegraphics[width=\columnwidth]{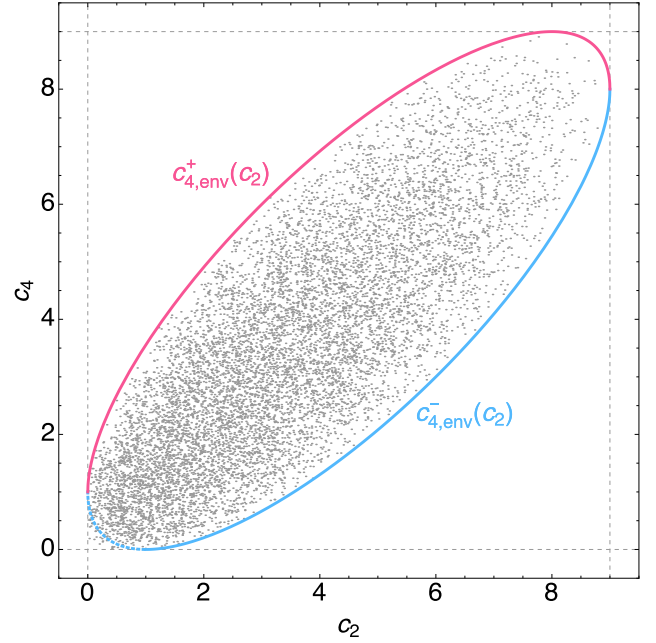}
\caption{
Loci of points satisfying criteria A-G, plotted in the $c_4$-$c_2$ plane.
The pink/blue lines are given by $c_{4,\mathrm{env}}^{\pm}(c_2)$.
}
\label{fig:c4c2surface}
\end{figure}

\begin{align}
c_3^2 &> (c_2 + c_4 - 1)^2,\,\,\mathbf{[A]}\nonumber\\
c_3^2 &< \frac{32 c_4 c_2}{9},\,\,\mathbf{[G]}\nonumber\\
\implies (c_2 + c_4 - 1)^2 &< \frac{32 c_4 c_2}{9},
\end{align}

where the last line may now be re-expressed as

\begin{align}
&\Big( c_4 - \frac{1}{9} \big( 9 + 7c_2 - 4\sqrt{2 c_2} \sqrt{9-c_2} \big) \Big) \nonumber\\
\times &\Big( c_4 - \frac{1}{9} \big( 9 + 7c_2 + 4\sqrt{2 c_2} \sqrt{9-c_2} \big) \Big) < 0.
\end{align}

The above requires that $c_4$ is greater than one of the radicals but always 
less than the other. Therefore, we have two valid envelope functions and we 
define the first one as:

\begin{align}
c_4 > c_{4,\mathrm{env}}^{-}(c_2) = \frac{1}{9} \big( 9 + 7c_2 - 4\sqrt{2 c_2} \sqrt{9-c_2} \big), \\
c_4 < c_{4,\mathrm{env}}^{+}(c_2) = \frac{1}{9} \big( 9 + 7c_2 + 4\sqrt{2 c_2} \sqrt{9-c_2} \big).
\end{align}

These two functions envelope the loci of points simulated earlier. This is 
apparent from Fig.~\ref{fig:c4c2surface}, where the pink line denotes 
$c_{4,\mathrm{env}}^{+}(c_2)$ and the blue line $c_{4,\mathrm{env}}^{-}(c_2)$. 
We plot the $c_{4,\mathrm{env}}^{-}(c_2)$ function from the upper-right 
intersection down to hitting the $c_4$ boundary condition of criterion F. 
Extending the line further back, shown as a dotted line, does not provide a 
physical bound on the points, although we note that only a very small fraction 
of points are excluded by doing so. In any case, the objective here is merely to
determine the upper limit on $c_4$.

We may use the $c_{4,\mathrm{env}}^{+}(c_2)$ function to find the maximum 
bound on $c_4$. Maximizing the $c_{4,\mathrm{env}}^{+}(c_2)$ function 
by differentiation, we find the curve is maximized at $c_2=8$, corresponding to
$c_4 = 9$. This provides the last limit needed to define the cuboid containing 
all loci as tightly as possible:

\begin{align}
0 < c_2 < 9,\nonumber\\
-8-6\sqrt{2} < c_3 < 0,\nonumber\\
0 < c_4 < 9.
\end{align}

\subsection{$c_4$-$c_3$ plane}
\label{sub:c4c3plane}

\begin{figure}
\includegraphics[width=\columnwidth]{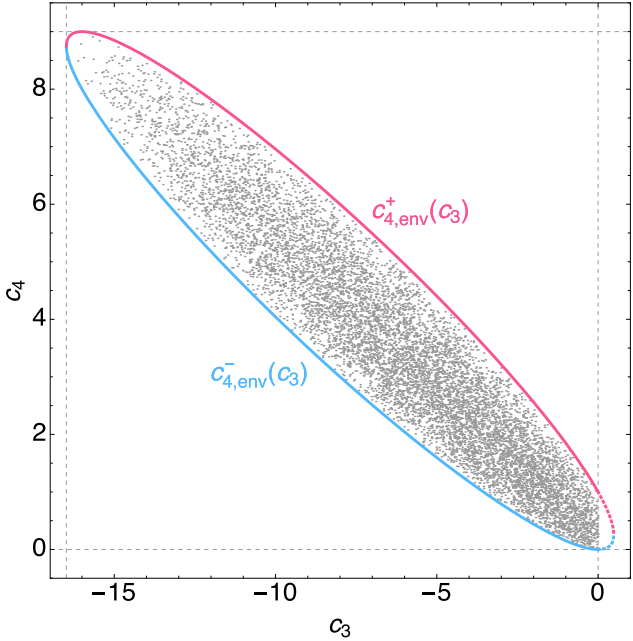}
\caption{
Loci of points satisfying criteria A-G, plotted in the $c_4$-$c_3$ plane.
The pink/blue lines are given by $c_{4,\mathrm{env}}^{\pm}(c_3)$.
}
\label{fig:c4c3surface}
\end{figure}

Although we have now derived all of the LDC bounds, for the sake of 
completeness we here consider envelope functions bounding the $c_4$-$c_2$ plane, 
as shown in Fig.~\ref{fig:c4c3surface}.

The pink and blue lines shown in Fig.~\ref{fig:c4c3surface} nearly perfectly
envelope the loci of allowed points, with the exception of criterion 
E ($c_3<0$) truncating a small fraction of points. As before, these functions
may be derived from the following:

\begin{align}
c_2 &< 1 - c_3 - c_4,\,\,\mathbf{[A]}\nonumber\\
c_2 &> \frac{9 c_3^2}{32 c_4},\,\,\mathbf{[G]}\nonumber\\
\implies \frac{9 c_3^2}{32 c_4} &< 1 - c_3 - c_4,
\end{align}

which may be re-expressed as

\begin{align}
32 &\Big( c_4 - \frac{1}{8}\big( 4 - 4c_3 - \sqrt{2}\sqrt{8-16c_3-c_3^2} \big) \Big) \nonumber\\
\times&\Big( c_4 - \frac{1}{8}\big( 4 - 4c_3 + \sqrt{2}\sqrt{8-16c_3-c_3^2} \big) \Big) < 0.
\end{align}

Following the lines of argument used before, this provides the envelope 
functions:

\begin{align}
c_4 > c_{4,\mathrm{env}}^{-}(c_3) = \frac{1}{8}\big( 4 - 4c_3 - \sqrt{2}\sqrt{8-16c_3-c_3^2} \big), \\
c_4 < c_{4,\mathrm{env}}^{+}(c_3) = \frac{1}{8}\big( 4 - 4c_3 + \sqrt{2}\sqrt{8-16c_3-c_3^2} \big).
\end{align}

We note that the two functions meet when $\sqrt{8-16c_3-c_3^2}=0$, occurring
at $c_3 = -8 + 6 \sqrt{2} = 0.4853...$. However, criterion G truncates these
functions, albeit only a small fraction of the envelope.

%% file: criteria_tests.tex

\subsection{Overview}

Starting from three physically imposed conditions for the intensity profile of a
limb darkened star, we have derived seven criteria which bound the three LDCs
parameterizing the \citet{sing:2009} limb-darkening law. In this section, we
test the validity of these criteria in terms of (i) completeness and (ii) 
validity. We define these terms as follows:

\begin{itemize}
\item[{\tiny$\blacksquare$}] \textbf{Completeness:} A fully complete set of 
criteria require that the loci of points for which the physical conditions are
met also satisfy our seven criteria.
\item[{\tiny$\blacksquare$}] \textbf{Validity:} A fully valid set of criteria 
requires that that the loci of points which meet our seven criteria 
never break the two physical conditions.
\end{itemize}

These two tests can be thought of in the following way. Incomplete cases imply 
that our criteria are overly-conservative, cropping some parts of physically 
plausible parameter space. Invalid cases imply that our criteria are
overly-optimistic, erroneously predicting that some parts of parameter space are
physically plausible.

\subsection{Completeness Tests}

We perform our tests via numerical Monte Carlo simulation. We begin by
drawing a uniform random point in the parameter space $\{c_2,c_3,c_4\}$. The 
expected upper and lower bounds on these terms were calculated earlier in 
\S\ref{sub:c4c2plane}. We begin by using these bounds except that we use the 
original $c_4>-1$ constraint rather than the modified criterion F. We then
take this cuboid and double the lengths of each side such that we consider
the ranges: $-9/2<c_2<27/2$, $-3(4+3\sqrt{2})<c_3<(4+3\sqrt{2})$ and
$-6<c_4<14$. These adjustments are made to ensure that we explore the full
range of physically allowed LDCs during this test.

A sample point is drawn from this cuboid and then tested as to whether the 
physical conditions \I, \II\ and \III\ are met. In practice, we accomplish this 
by computing $10^3$ points along the functions $I(\mu)$ and 
$\partial I(\mu)/\partial\mu$ varying $\mu$ from $0$ to $1$ in equal, linear 
steps. For \III\ we simply test if $c_3<0$, since this condition only applies
at the location $\mu\to0$. If the physical conditions are not met, then we 
generate a new trial point. If they are met, then we proceed to test if 
the analytic criteria A to G are satisfied for this accepted point.

In total, we repeat this process until $10^4$ accepted points are found, 
requiring $\sim10^7$ simulations in total. Although the number of simulations 
may appear modest, we note that at each realization we must numerically compute 
the functions $I(\mu)$ and $\partial I(\mu)/\partial\mu$ at $10^3$ locations, 
which takes substantial computational overhead ($\sim30$\,s per simulation).

We find that $95.3$\% of the accepted points also satisfy criteria A to G, or
a completeness of $>95$\%. This indicates that our seven criteria are slightly
overly-conservative, cropping $\sim5$\% of the physically permissible 
LDCs. If we use the unmodified versions of criteria F and G (see 
\S\ref{sub:criterionF} and \S\ref{sub:criterionG}) and repeat the exercise,
we find that $100$\% of the physically valid points satisfy the criteria.
However, as discussed in \S\ref{sec:transformations}, this now yields an
asymmetric loci of allowed LDCs, impeding efforts to find an efficient sampling
algorithm.

\subsection{Validity Tests}
\label{sub:validitytests}

In an analogous approach to the previous tests, we begin by drawing uniform 
random samples in $\{c_2,c_3,c_4\}$ as before, except that we know constrain the 
cuboid to the specific bounds derived in \S\ref{sub:c4c2plane} (including 
$c_4>0$). We then test whether the seven analytic criteria are satisfied or not 
and if so consider the point to be accepted. We continue until $10^6$ accepted 
points are found, which we found required $101,163,869$ trials. Since we used 
the tightest bounding cuboid possible here, this reveals that the most efficient
sampling possible without transforming the $\{c_2,c_3,c_4\}$ LDCs would be
just under 1\%. Whilst one could proceed in this way, applying an 
acceptance/rejection test at each realization, uniform sampling would reject
over 99\% of realizations, making such an approach highly inefficient.

We next test whether each of these accepted points satisfies the physical
conditions \I, \II\ and \III. As before, this is done by evaluating the 
functions $I(\mu)$ and $\partial I(\mu)/\partial\mu$ at $10^3$ evenly spaced 
locations.

From these tests, we find that 100\% of the $10^6$ accepted points satisfy the
physical criteria. Therefore, the seven criteria are fully valid and drawing
a point which satisfies them is guaranteed to always satisfy the physical 
conditions \I, \II\ and \III.

These tests therefore confirm that our criteria have a very high completeness
and perfect validity. We therefore proceed with confidence that
they provide a suitable set of constraints to evaluate the range of physically
plausible LDCs.

\begin{figure}
\includegraphics[width=\columnwidth]{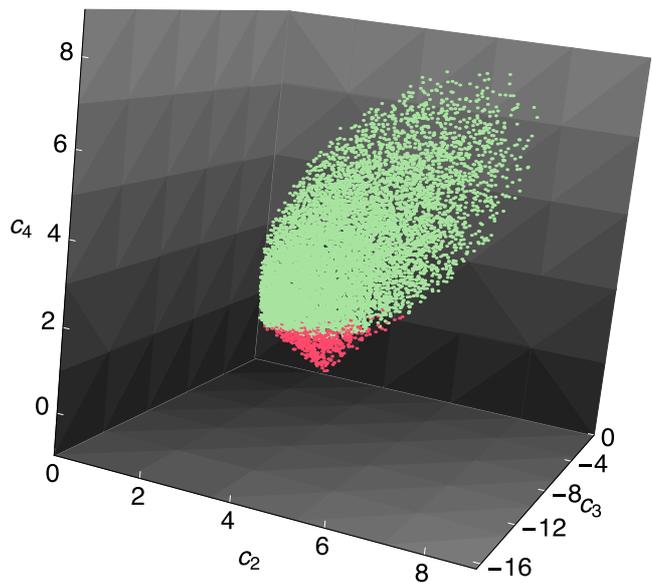}
\caption{
3D plot of the allowed LDCs in the original parameter space: $\{c_2,c_3,c_4\}$. 
All of the plotted points satisfy the physical conditions \I, \II\ and \III,
as well as the unmodified criteria A-G. The green points also satisfy the 
modified criteria A-G (whereas the red do not), chosen to yield a more symmetric 
loci of points and comprising $>95$\% of the volume.
}
\label{fig:c3Dplot}
\end{figure}

%% file: transformations.tex

\subsection{Overview}

When plotted in the original $\mathbf{c}=\{c_2,c_3,c_4\}$ parameter space 
(Fig.~\ref{fig:c3Dplot}), the loci of physically allowed LDCs resemble a 
rotated, tilted ellipse of thin but finite width with a symmetric protrusion 
running along the semimajor axis. This complex morphology cannot be easily
sampled from and if one wished to draw a set of LDCs from a uniform prior, there
would be no alternative except to draw from the full cuboid and perform
an acceptance/rejection test using our seven analytic criteria. As evident from
Fig.~\ref{fig:c3Dplot}, the volume of allowed points is much less than the 
bounding cuboid volume, meaning that such a procedure would be highly 
inefficient. Numerical tests reveal that the efficiency of such a procedure is
just under 1\%, making any algorithms using this method very wasteful.

We are therefore motivated to try and transform the geometry of the accepted
loci of points into a more regular shape that we can sample from efficiently.
We previously did this in the 2D case of quadratic limb darkening in 
\citet{LD:2013}, but the transformative geometry required here is not only
more complex but also includes an extra dimension.

\subsection{Rescaled LDCs}

We begin by noting that the envelope functions shown in 
Fig.~\ref{fig:c3c2surface}-\ref{fig:c4c3surface} provide an excellent 
description for the bounding region of allowed LDCs. Despite some small 
exceptions, we are motivated to exclusively use these simple envelopes 
rather than the full criteria since i) they provide a nearly perfect description 
of the loci ii) the envelopes are symmetric functions derived from quadratic 
forms iii) all of the envelopes come from criteria A and G alone. In practice, 
criterion F is also necessary to remove a duplicate set of solutions.

We therefore proceed to only consider the region contained by criteria A, F and 
G, which we denote as the simplified region. This simplification means that the
bounding cuboid is slightly modified to:

\begin{align}
0 < c_2 < 9,\nonumber\\
-8-6\sqrt{2} < c_3 < -8+6\sqrt{2},\nonumber\\
0 < c_4 < 9.
\end{align}

As a first transformation, we re-scale the axes into a unitary cube by the use 
of $d_i$ terms, defined as:

\begin{align}
d_2 &= c_2/9, \nonumber\\
d_3 &= \frac{6\sqrt{2} - 8 - c_3}{12\sqrt{2}}, \nonumber\\
d_4 &= c_4/9.
\end{align}

\subsection{Righting the Allowed Region}

We next note that the loci (in $\mathbf{d}$ parameter space) resembles an
elliptic thin disk with the semimajor axis pointing along the unit vector 
$\{1,1,1\}$. Further, the disk appears rotated about this unit vector, with
respect to the axes of the reference frame. We decided to try to right the 
volume by performing a clockwise rotation of $(\pi/3)$\,radians about the unit 
vector ($\mathbf{M}_1$). We then perform a further rotation which relocates the 
unit vector $\{1,1,1\}$ to $\{1,1,0\}$ ($\mathbf{M}_2$), followed by a third 
rotation relocating $\{1,1,0\}$ to $\{1,0,0\}$ ($\mathbf{M}_3$). The total 
rotation matrix applied is described by:

\begin{align}
{\bf e} &= {\bf M_3}.{\bf M_2}.{\bf M_1}.{\bf d},\nonumber\\
{\bf e} &= {\bf M}.{\bf d},
\end{align}

where

\begin{eqnarray}
{\bf M} =\left(\begin{matrix}
\frac{1}{\sqrt{3}} & \frac{1}{\sqrt{3}} & \frac{1}{\sqrt{3}} \cr
-\frac{1}{\sqrt{2}} & 0 & \frac{1}{\sqrt{2}} \cr
\frac{1}{\sqrt{6}} & -\frac{\sqrt{2}}{\sqrt{3}} & \frac{1}{\sqrt{6}} \cr \end{matrix}\right).
\label{eqn:Mrot}
\end{eqnarray}

Applying these transformations gives a new-coordinate system of:

\begin{align}
e_2 &= \frac{ 36 - 24\sqrt{2} + 8c_2 - 3\sqrt{2}c_3 + 8c_4 }{ 72\sqrt{3} },\\
e_3 &= \frac{ c_4 - c_2 }{ 9\sqrt{2} },\\
e_4 &= \frac{ 24 - 18\sqrt{2} + 2\sqrt{2}c_2 + 3c_3 + 2\sqrt{2}c_4 }{ 36\sqrt{3} }.
\label{eqn:espace}
\end{align}

\begin{figure}
\includegraphics[width=\columnwidth]{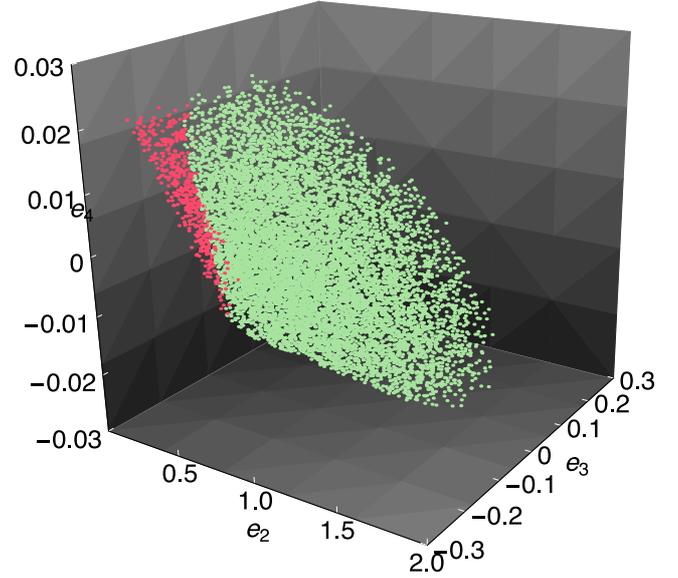}
\caption{
Same as Fig.~\ref{fig:c3Dplot}, except the coordinates have been transformed
from $\{c_2,c_3,c_4\}\to\{e_2,e_3,e_4\}$. In this parametrization, we observe
that the volume of green points resembles a cone.
}
\label{fig:e3Dplot}
\end{figure}

After visually inspecting the loci in this transformed parameter space (as
shown in Fig.~\ref{fig:e3Dplot}), we note that the allowed region now resembles 
a cone. Motivated by this observation, we proceed to transform this shape into a 
cone of symmetric proportions and with principal axes aligned to the transformed 
frame.

\subsection{The Conal Region}

As with the $\mathbf{c}$ parameter space, our first step is to re-scale the
axis into a cuboid with unit lengths, requiring us to first compute the extrema
in $\mathbf{e}$ space.

The extrema of $e_3(c_2,c_4)$ occur when $(c_4-c_2)$ is maximized/minimized,
as evident from Equation~\ref{eqn:espace}. This can be considered further by 
studying our earlier illustration in Fig.~\ref{fig:c4c2surface}. This can be 
found by maximizing the function $(c_4^{\pm}-c_2)$ with respect to $c_2$, which 
reveals the extrema is $-3<(c_4-c_2)<+3$. We are therefore able to show that 
$-\frac{1}{3\sqrt{2}}<e_3<\frac{1}{3\sqrt{2}}$. This can also be achieved by
maximizing/minimizing the $e_3$ expression using the additional constraints of
criteria A, F and G. Repeating for the other terms we define a new boundary box
of

\begin{align}
\frac{ 3\sqrt{3} - 2\sqrt{6} }{ 18 } < &e_2 < \frac{ 2 (3+\sqrt{2}) }{ 3 \sqrt{3} },\\
-\Bigg(\frac{1}{3\sqrt{2}}\Bigg) < &e_3 < \Bigg(\frac{1}{3\sqrt{2}}\Bigg),\\
-\Bigg(\frac{ 3\sqrt{2} - 4 }{ 6\sqrt{3} }\Bigg) < &e_4 < \Bigg(\frac{ 3\sqrt{2} - 4 }{ 6\sqrt{3} }\Bigg).
\end{align}

We choose to re-scale the $e$-parameter space into a unit vector cube via:

\begin{align}
f_2 &= \frac{ e_2 - \frac{ 3\sqrt{3} - 2\sqrt{6} }{ 18 } }{ \frac{\sqrt{2}}{\sqrt{3}} + \frac{\sqrt{3}}{2} } ,\\
f_3 &= \frac{ 3 e_3 }{ \sqrt{2} },\\
f_4 &= \frac{ e_4 + \frac{ 3\sqrt{2} - 4 }{ 6\sqrt{3} } }{ \frac{ 6\sqrt{2} - 8 }{ 6\sqrt{3} } }.
\end{align}

At this point, we now choose to rotate the conic section by $\pi/4$ radians
in a clockwise sense around the $f_3$-axis, so that the cone's apex is located
at the origin and the cone points along the $e_3$-axis. We accomplish this using
an additional change of variables:

\begin{align}
g_2 &= \frac{f_2-f_4}{2},\\
g_3 &= f_3,\\
g_4 &= \frac{f_2+f_4}{\sqrt{2}},
\end{align}

where we have additionally normalized $g_2$ by a factor of $\sqrt{2}$ to allow
the loci to be symmetric on the $g_2$-$g_3$ plane.

\begin{figure}
\includegraphics[width=\columnwidth]{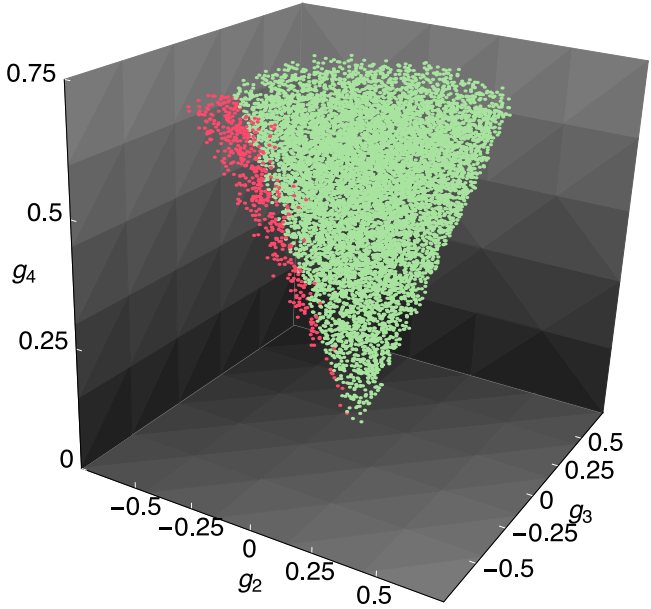}
\caption{
Same as Fig.~\ref{fig:c3Dplot}, except the coordinates have been transformed
from $\{c_2,c_3,c_4\}\to\{g_2,g_3,g_4\}$. In this parametrization, the green
points are well described by a cone of radius, $R=1/2$ and height, 
$H=(-4 + 10\sqrt{2}/3)$.
}
\label{fig:g3Dplot}
\end{figure}

In this frame, our cone now has an apex at zero, with a height of
$H = (\frac{10\sqrt{2}}{3} - 4)$ and a radius of $R = 1/\sqrt{2}$, as shown in
Fig.~\ref{fig:g3Dplot}. Writing out the $g$ terms relative to the original 
$c_i$ coefficients, we have

\begin{align}
g_2 &= \frac{1}{72\sqrt{2}} \Big( (6\sqrt{2}-56)c_2 + (-6\sqrt{2}-45)c_3 + (6\sqrt{2}-56)c_4 \Big),\\
g_3 &= \frac{1}{6} \Big( c_4 - c_2 \Big),\\
g_4 &= \frac{1}{72} \Big( (42\sqrt{2}-8)c_2 + (30\sqrt{2}+9)c_3 + (42\sqrt{2}-8)c_4 \Big).
\end{align}

For which the inverse relations are

\begin{align}
c_2 &= \Big( \frac{3}{2} + 5\sqrt{2} \Big) g_2 - 3 g_3 + \Big( 1 + \frac{ 15 }{ 2\sqrt{2} } \Big),\\
c_3 &= \Big( \frac{8}{3} - 14\sqrt{2} \Big) g_2 + \Big( 2 - \frac{28\sqrt{2}}{3} \Big) g_4,\\
c_4 &= \Big( \frac{3}{2} + 5\sqrt{2} \Big) g_2 + 3 g_3 + \Big( 1 + \frac{15}{2\sqrt{2}} \Big) g_4.
\end{align}

\subsection{Sampling from the Conal Region}

The samples shown in Fig.~\ref{fig:g3Dplot} appear consistent with points
uniformly drawn from within the volume of a cone. We here describe the
mathematical formalism by which one can compute such samples.

Samples may be drawn from a cone by first considering how to draw samples
uniformly from within a circle. This well-known problem can be tackled by
using polar coordinates and drawing a random polar angle $\theta$ in the range
$0$-$2\pi$\,rad and a random radius $r$ from a triangular distribution
between $0$ and $\rho$, where $\rho$ is the full radius. We now note that the
radius varies as a function of height, $h$, along the cone, such that 
$\rho(h) = R h/H$. Finally, $h$ is drawn from a quadratic power-law
distribution from $0$ to $H$ (the full height), since the area of a circle 
increases as $\rho^2$. Drawing a random uniform variate for $\alpha_{\theta}$,
$\alpha_h$ and $\alpha_r$ between 0 and 1, the polar angles, height and radius
of a point uniformly drawn from within the cone may be expressed as

\begin{align}
\theta &= 2 \pi \alpha_{\theta},\\
h &= H \alpha_h^{1/3},\\
r &= \frac{ R h \sqrt{\alpha_r} }{ H }.
\end{align}

Converting these into Cartesian elements, we have

\begin{align}
g_2 &= r \sin \theta,\\
g_3 &= r \cos \theta,\\
g_4 &= h.
\end{align}

Or more explicitly:

\begin{align}
g_2 &= R \alpha_h^{1/3} \alpha_r^{1/2} \sin ( 2 \pi \alpha_{\theta} ),\\
g_3 &= R \alpha_h^{1/3} \alpha_r^{1/2} \cos ( 2 \pi \alpha_{\theta} ),\\
g_4 &= H \alpha_h^{1/3} .
\end{align}

We may also express the $c_i$ coefficients in terms of the uniform random
variates, $\alpha_i$:

\begin{align}
c_2 =& \frac{\alpha_h^{1/3}}{12} \Bigg( 28 (9-5\sqrt{2}) \nonumber\\
\qquad& + 3 \alpha_r^{1/2} \Big( -6\cos(2\pi\alpha_{\theta}) 
+ (3+10\sqrt{2}\sin(2\pi\alpha_{\theta}) \Big) \Bigg),\\
c_3 =& \frac{\alpha_h^{1/3}}{9} \Bigg( -632 + 396 \sqrt{2} \nonumber\\
\qquad& + 3\alpha_r^{1/2}(4-21\sqrt{2})\sin(2\pi\alpha_{\theta}) \Bigg) ,\\
c_4 =& \frac{\alpha_h^{1/3}}{12} \Bigg( 28 (9-5\sqrt{2}) \nonumber\\
\qquad& + 3 \alpha_r^{1/2} \Big( 6\cos(2\pi\alpha_{\theta}) 
+ (3+10\sqrt{2}\sin(2\pi\alpha_{\theta}) \Big) \Bigg).
\label{eqn:alpha}
\end{align}

One may now work in the $\boldalpha$ parameter space, drawing samples from 
within a unit cube (see Fig.~\ref{fig:alpha3Dplot}) and then converting into a 
physically plausible set of LDCs using the above.

A unique set of inverse relations can be defined by use of an arc tangent
accounting for the quadrant of the radical and the use of a floor function.
We have written \python\ and \fortran\ code, called \LDC, to perform
these functions, which is publicly available at \link.

\subsection{Testing Samples Drawn from the Conal Region}

We have provided no formal proof that the loci of points in $g$-space is bound
by a cone, nor do we explicitly claim so. We merely observe that the morphology 
of the loci most closely resembles this shape, from which it is possible to 
easily draw uniform samples. We here provide two simple tests demonstrating that 
sampling from the conal region provides an excellent set of physical LDCs.

First, we generated $10^6$ uniform random points from within the cone and 
tested whether they satisfied the physical conditions \I, \II\ and \III. 
Using $R=1/2$ and $H=(-4 + 10\sqrt{2}/3)$, we find that 97.3\% of the conal
region is physical, or equivalently, points sampled from this region have a 
validity of 97.3\%. Similarly, using the sample of $10^4$ valid points generated 
earlier in \S\ref{sub:validitytests}, we find a completeness of 94.4\%. 
Therefore, points samples from the conal region crop $\sim5$\% of the allowed 
parameter space.

\begin{figure}
\includegraphics[width=\columnwidth]{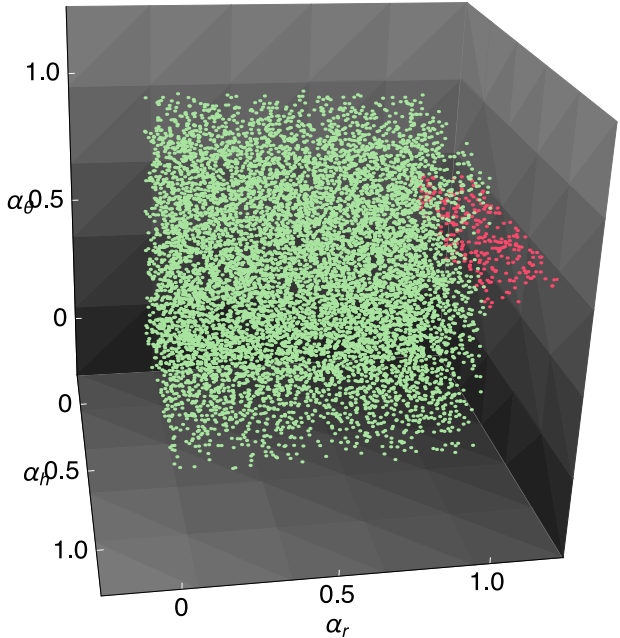}
\caption{
Same as Fig.~\ref{fig:c3Dplot}, except the coordinates have been transformed
from $\{c_2,c_3,c_4\}\to\{\alpha_h,\alpha_r,\alpha_{\theta}\}$. In this 
parametrization, the green points are nearly uniformly distributed within a
unit cube. One may therefore uniformly sample from the cube in 
$\boldalpha$-space and then transform back to $\mathbf{c}$-space to efficiently
sample physical LDCs.
}
\label{fig:alpha3Dplot}
\end{figure}

\begin{figure}
\includegraphics[width=\columnwidth]{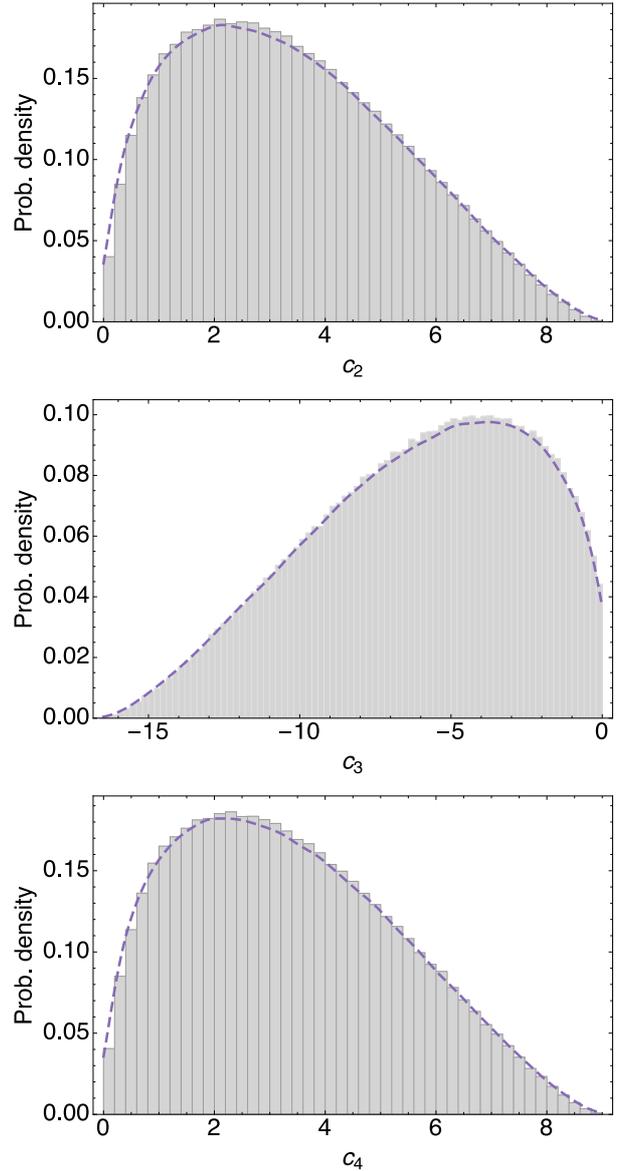}
\caption{
Dashed, purple lines depict smoothed histograms of $10^6$ LDCs generated by 
uniformly sampling from the conal region and then transforming from $\boldalpha 
\to \mathbf{c}$ parameter space. Grey histograms depict $10^6$ LDCs drawn from a 
random uniform prior in $\mathbf{c}$ parameter space which also satisfy the 
seven analytic criteria. The close agreement demonstrates the effectiveness of 
directly sampling from the conal region to draw physical LDCs.
}
\label{fig:chistos}
\end{figure}

Aside from validity and completeness, we also consider the distribution of LDCs
generated from sampling the conal region. We find that uniform samples from the
$\mathbf{g}$-cone (or equivalently uniform points in the $\boldalpha$-cube)
yield $\{c_2,c_3,c_4\}$ LDCs closely matching the distribution which would
result from uniform sampling in $\mathbf{c}$-space with a simple 
acceptance/rejection of criteria A-G. This is evident in 
Fig.~\ref{fig:chistos}, where we compare the nearly identical distributions 
from these two approaches. One can also see in Fig.~\ref{fig:alpha3Dplot}, 
that the valid samples plotted in $\boldalpha$-space (which were initially drawn 
uniformly in $\mathbf{c}$-space) provide an approximately uniform set of points 
within the unit cube.

%% file: discussion.tex

In this work, we have presented a set of seven analytic criteria which may be
used to assess the physicality of LDCs associated with the \citet{sing:2009}
three-parameter non-linear limb-darkening law. We imposed simple conditions that 
the flux is everywhere positive, monotonically decreases from centre to limb and 
has a negative curl at the limb. Through numerical testing, we have shown that 
points naively sampled with a simple accept/reject algorithm applied to our 
criteria are always physically valid. Additionally, over 95\% of the physically 
allowed loci of LDCs (found through brute force numerical exploration) satisfy 
the seven criteria, demonstrating a very high completeness. Using an unmodified
set of criteria which retains the asymmetries present in the loci of allowed
LDCs, the completeness is 100\%.

Armed with these criteria, we have re-parametrized the LDCs such that the loci 
of allowed points morphologically resemble a regular geometric shape, 
specifically a cone. We have shown that uniformly sampling points from the conal
region in the re-parametrized space yields physically plausible and
uniformly distributed LDCs to high accuracy. Specifically, we find a validity
of 97.4\% and a completeness of 94.4\%. 

Sampling from the conal region may be achieved by drawing a uniform random 
variate in the transformed space $\{\alpha_h,\alpha_r,\alpha_{\theta}\}$, for 
which the relational expressions to the original $\{c_2,c_3,c_4\}$ LDCs are 
provided in Equation~\ref{eqn:alpha}. We also provide public code (\LDC) in 
\python\ and \fortran\ to perform both the forward and inverse calculation
between the parametrizations (\link).

Our work provides, for the first time, a practical and efficient framework for 
fitting astronomical data affected by limb-darkening with a law supporting three
degrees of freedom. Until now, one had to limit oneself to efficient sampling 
of a two-parameter limb-darkening law \citep{LD:2013} and go without the major
improvement in accuracy provided by a three-parameter law, such as that of
\citet{sing:2009}. Alternatively, one would have had to explore and marginalize
over unphysical combinations of LDCs (which we estimate would occur for at 
least 99.9\% of naively sampled points) or numerically test the physicality
of each realization of LDCs (again with an overhead of rejecting the vast
majority of points). In any case, we argue that our solution provides major
advantages and enables the community to practically fit more complex 
limb-darkening profiles for the first time.

The only published grids of theoretical LDCs using the \citet{sing:2009} law
comes from \citet{sing:2010}. With the Kepler bandpass, we find that 99.6\% of 
the \citet{sing:2010} tabulated points satisfy physical conditions \I, \II\ and 
\III. Further more, 97.7\% of these tabulated points reproduce an $\boldalpha$
transformed LDC within the unit cube. These values again demonstrate that
the $\boldalpha$ parametrization can be practically used to explore the
physically allowed LDCs. Since we now have an efficient strategy to fit LDCs,
there is the potential to verify the predictions made from theoretical models
by the study of high signal-to-noise transits in the future. 

In some applications, having ``only'' $97.3$\% of the LDCs being physically 
valid may be insufficient and one may wish to ensure 100\% validity. Since 
the seven analytic criteria guarantee 100\% validity, one may draw a set of LDCs 
using the $\boldalpha$ parametrization and then test if this realization 
satisfies the seven criteria. In practice, criteria B and D are never
violated by points sampled from the conal region and thus it is only necessary
to test five criteria. This approach enables a guaranteed physically plausible 
set of LDCs at minimal computational expense. To aid the community, our code 
\LDC\ can perform this test (\link).

Whilst the quadratic law (and other two parameter laws) will likely remain 
suitable for many studies, the analysis of high precision data increasingly 
demands a more sophisticated treatment of limb-darkening to avoid this
issue becoming a bottleneck in obtainable accuracy \citep{epinoza:2015}. By 
freely fitting high-precision data with our $\boldalpha$ parametrization of the
\citet{sing:2009} limb-darkening, one can have greater confidence that the 
parameters of interest are marginalized exclusively over the physically 
plausible parameter space and limb-darkening is modelled in a manner more 
consistent with simulations from modern stellar atmosphere models. We also note 
that informative priors on our $\boldalpha$ parametrization may be used as 
well, in cases where one has strong belief in the results of stellar atmosphere 
models and the star is well-characterized already, or alternatively from 
previous posteriors derived from freely fitting the LDCs. For either
informative or uninformative sampling, the $\boldalpha$ parametrization offers
an efficient and physically sound pathway to exploring parameter space when
modelling limb-darkening under the three-parameter law.